\documentclass{aastex}
\usepackage{spr-astr-addons}
\usepackage{url}
\usepackage{amsmath}
\usepackage{amsfonts}
\usepackage{amssymb}
\usepackage{graphicx}
\usepackage{subfigure}

\RequirePackage{color}

\begin{document}

\title{Equivalence Principle (EP) and Solar System Constraints on $R(1\pm \epsilon \ln({R \over R_c}))$ model of Gravity}
\slugcomment{Not to appear in Nonlearned J., 45.}
\shorttitle{Short article title}

\shortauthors{Kh. Saaidi et al.}
\author{$^a$Kh. Saaidi\altaffilmark{1}}\and\author{$^b$A. Aghamohammadi\altaffilmark{2}}
 \affil{$^a$Islamic Azad University, Sanandaj branch,  Sanandaj, Iran.}\and\affil{$^b$Plasma Physics Research Center, Science and Research
Branch Islamic Azad University of Tehran,
   Iran  }
   \altaffiltext{1}{ksaaidi@uok.ac.ir.}
\altaffiltext{2}{agha35484@yahoo.com.}


\begin{abstract}
Experiments on the violation of equivalence principle (EP) and solar system give a number of constraints in which any modified gravity model
   must satisfy them. We study these constraints on a kind of $f(R)$ gravity as $f(R) = R(1\pm \epsilon \ln({R \over R_c}))$. For this investigation we use of chameleon mechanism and show that a spherically body has thin-shell in this model. So that we  obtain an effective
   coupling of the fifth force which is suppressed through a chameleon mechanism. Also, we obtain $\gamma_{PPN} = 1 \pm 1.13 \times10^{-5}$  which is agreement with experiment results. At last, we show that for $R_c \thickapprox \rho_c$ this model is consistent with EP, thin shell condition and
   fifth force of chameleon mechanism for $\epsilon \backsimeq 10^{-14}$.

\end{abstract}


\section{Introductions}
In the recent decade,   the acceleration of the universe expansion was discovered and is still a deep mystery, (for review see e.g (\cite{Brax}, \cite{3},  \cite{4},\cite{5} )).  Two approaches  have been introduced for interpret  the present acceleration. One can either introduce  an unknown form of energy, dubbed dark energy, that  suggest about  \%70  of energy density of the present universe is composed by it or modify the behavior of gravity at cosmological distances. In the first approach,  the most relevant candidate for the role of dark energy is Einstein's cosmological constant, which interpret the cosmic expansion in the $\Lambda CDM$ model (\cite{Bisabr}, \cite{7}, \cite{8}, \cite{9}) but in order to overcome its intrinsic  shortcomings associated with the energy scale, several alternative models such as quintessence, etc have been proposed (\cite{10}). Most of these models have the common feature to introduce new sources  in to the cosmological dynamics, but these models have the cosmological constant problem, the coincidence problem and the value of equation of state. On the other hand, in the second approach, various attempts to modify gravity have been presented(\cite{11},\cite{12} ,\cite{14},\cite{15}), (\cite{18}, \cite{19}, \cite{20}) (\cite{21},\cite{Ali}), (\cite{16}). Among different approaches, there are modified gravity models  so called $f(R)$, namely  (we replace the Ricci scalar $R$ in the Einstein-Hilbert lagrangian density to some function $f(R)$), which do not seem to introduce new type of matter and can lead to late time acceleration. In fact, these theories can be reformulated in terms of scalar tensor theories with a stabilized coupling of the additional scalar degree of freedom to matter. As theories of dark energy,  they suffer from the usual problems and are also potentially ruled out by gravitational tests of Newton's law. Among all cosmologically viable $f(R)$ theories  there is still an important issue to be pursued, they must be probed at solar system scale. This claim was based on the fact that $f(R)$ theories are equivalent to Brans-Dicke theory $\omega=0$,  while observations set the constraint $\omega>40000$ (\cite{22}), and the post-Newtonian parameter satisfies $\gamma_{PPN}=\frac{1}{2}$
instead of being equal to unity as needed by observations(\cite{a1}, \cite{a2}, \cite{Bisabr}). The only way-out for these models is that
behave as chameleon theories (\cite{Khoury}), i.e. evolve a field  mass dependent on the  local matter density
(\cite{24},  \cite{26}) (\cite{25}). The scalar field mediates a fifth force  which is suppressed in the laboratory and in interactions between large bodies such as planets, but which may be detectable between small test masses in space (\cite{27}). In the massive bodies, the fifth force is attenuated as the chameleon is trapped inside  very massive bodies (the sun for instance) (\cite{10}).  It has argued that the existence of thin shell is usually adequate to salvage $f(R)$ gravity models (\cite{27}).  Meanwhile, in Ref. (\cite{16}) the authors derived the conditions under which a successful sequence of radiation, matter and accelerated epochs can be realized.
In addition the stability conditions $f_{,R}>0$ and $f_{,R,R}>0$ are required to avoid ghosts and tachyon
for $R\geq R_1$, where $R_1$ is the Ricci scalar de-Sitter point(\cite{28}). There are viable $f(R)$ models that can satisfy both cosmological constraints and stability conditions(\cite{su1}, \cite{29}, \cite{30}, \cite{31}, \cite{32}).\\
   In this work  we have used  chameleon mechanism to place constraints as a local experiments on a viable $f(R)$ gravity model, that in the our previous work (\cite{1}), have been proposed. Whereas some proposal of $f(R)$ model of gravity could not consistent with astrophysical and
   experimental data, we want to find a model of $f(R)$ gravity which has been the most agreement with experimental data.  \\
The paper is organized as follows.  In first  section, we introduce a $f(R)$ model, then we apply the local gravity experiment constraints, as thin shell condition, equivalence principle, fifth force on viability of the model. Finally, the latter section is devoted to conclusions.
\section{Primary  Calculations}
In a static and spherically symmetric space time, with the metric
$$\bar{g}_{\mu\nu}=diag(1,-1,-\bar{r}^2,-\bar{r}^2\sin ^2\theta),$$
 the field  equation (\cite{Brax}) gives:
\begin{eqnarray}\label{9}
\frac{d^2 \phi}{{d\bar{r}}^2}+\frac{2}{\bar{r}}\frac{d\phi}{d\bar{r}}=\frac{dV_{eff}}{d\phi},
\end{eqnarray}
where $\bar{r}$ is the distance from the center of symmetry and
\begin{eqnarray}
V_{eff}(\phi)=V(\phi)+e^{\beta \phi}\bar{\rho}.\label{10}
\end{eqnarray}
Here $\bar{\rho}$ is the energy density in the Einstein frame, which is connected to the energy density $\rho$
in the Jordan frame via the relation $\bar{\rho}=e^{3\beta\phi}\rho$.
We assume that a spherically symmetric body with radius $\bar{r_c}$ has a constant energy density $\bar{\rho}_{in}$ inside body ($\bar{r}<\bar{r}_c$) and the energy density $\bar{\rho}_{out}$ outside the body ($\bar{r}>\bar{r}_c$). The mass $M_c$ and the gravitational potential $\Phi_c$ of the body with radius $\bar{r}_c$ are given by $M_c=\frac{4\pi}{3}\bar{r}^3_c\bar{\rho}$ and $\Phi_c=\frac{M_c}{8\pi\bar{r}_c}$, respectively.
For obtain $V_{eff}$, originally, we introduce a $f(R)$ model that can satisfy local gravity constraints as well as cosmological and stability condition which  we have proposed that in the our previous work as follows
\begin{equation}\label{11}
f(R)=R+R\ln {[\frac{R}{R_{c}}]}^{\mp\epsilon},
\end{equation}
where $R_c$ is positive constants and $\epsilon$ is a small and dimensionless constant. It is clear $f(R)\vert_{R=0}=0$,  on the flat space time and  in  the $\epsilon\ll 1 $, Eq. (\ref{11}) reduced to
\begin{equation}\label{12}
f(R)\backsimeq R({\frac{R}{R_c}})^{\mp\epsilon}.
\end{equation}
Also, in (\cite{Brax})
have been  given that
\begin{eqnarray}
V(\phi)=\frac{Rf'(R)-f(R)}{2{f'(R)}^2}.\label{5}
\end{eqnarray}
Then, with substituting (\ref{12}) into (\ref{5}) the function $V_{eff}(\phi)$, will become
\begin{equation}\label{13}
V_{eff}=\mp\frac{\epsilon R_c}{2}e^{2\beta\phi(1\pm\frac{1}{\epsilon})}+\bar{\rho}e^{\beta\phi}.
\end{equation}
It is obviously seen that the potential
$$V(\phi)=\mp\frac{\epsilon R_c}{2}e^{2\beta\phi(1\pm\frac{1}{\epsilon})}$$ satisfy the conditions of chameleon mechanism
 \begin{equation}\label{8a}
  \frac{dV}{d\phi}<0, \qquad \frac{d^2 V}{{d \phi}^2}>0,\qquad  \frac{d^3 V}{d^3\phi}<0.
  \end{equation}
Assuming $\phi\ll\epsilon\ll 1$, one can find the solution of $V'_{eff}=0$, with hypothesis
$\bar{\rho}\ll \frac{2R_c}{\epsilon}$ as
\begin{equation}\label{14}
\phi_{min}=[\frac{-\bar{\rho}+R_c}{2\beta R_c}] \epsilon.
\end{equation}
The effective potential $V_{eff}$ has two minima at the field value $\phi_{in}$ and $\phi_{out}$ according to (\ref{14}), established upon $\bar{\rho}_{in}$  and $\bar{\rho}_{out}$, respectively. Here, the $\phi_{in}$ correspond to  the region with a high density that gives rise to a heavy mass squared,
whereas the $\phi_{out}$ to the lower density region with a lighter mass. Generally, the  masses of scalar fields about these minima are given by $m_{in}^2=\frac{d^2V_{eff}(\phi_{in})}{{d\phi}^2}$ and   $m_{out}^2=\frac{d^2V_{eff}(\phi_{out})}{{d\phi}^2}$.
In this regard,  the near  of massive bodies with a heavy field mass, it is known that the spherically symmetric body has  a thin-shell under the chameleon mechanism, i.e.
\begin{equation}\label{15}
\frac{\Delta \bar{r}_c}{\bar{r}_c}=\frac{\phi_{out}-\phi_{in}}{6\beta \phi_c}\ll 1.
\end{equation}
Solution, equation (\ref{9}) with appropriate boundary conditions $\phi_{out}=\phi(r=\infty)$ gives the field profile of  the body($\bar{r}>\bar{r}_c$)
 \begin{equation}\label{16}
\phi(\bar{r})\simeq -\frac{\beta}{4\pi}\frac{3\Delta \bar{r}_c}{\bar{r}_c}\frac{M_ce^{-m_{out}(\bar{r}-\bar{r}_c)}}{\bar{r}}+\phi_{out}.
 \end{equation}
  \section{Thin-Shell Condition}
  In the chameleon mechanism,  the chameleon field is trapped inside massive bodies and its influence on the other bodies is alone caused by a thin-shell close to the surface of the body (\cite{Khoury}). Hence, by substituting (\ref{14}) and (\ref{16}) into the thin-shell  condition,(\ref{15}),  we have
  \begin{equation}\label{17}
\frac{\Delta \bar{r}_c}{\bar{r}_c}=[\frac{\bar{\rho}_{in}-\bar{\rho}_{out}}{12{\beta}^2\Phi_c R_c}]\epsilon.
\end{equation}
 Where $\bar\rho_{in}$ and $\bar\rho_{out}$ are energy densities inside and outside of the body in the Jordan frame. Note that $R_c$ is not very different from $\rho_{in}$. Since $ \rho_{out}\ll \rho_{in}\ll \frac{\Phi_c R_c}{\epsilon}$, also $m_{out}r\ll 1$, means that  the Compone wavelength $m_{out}^{-1}$ is very larger than Solar System scales, then  in the relation $\bar{r}=e^{-2\beta \phi}r $ we can apply  $\bar{r}\simeq (1-2\beta\phi+\cdots)\simeq r$. In the following we eliminate  the bar upon the quantity $r$. We will first discuss post-Newtonian solar- system constraints on the model (\ref{11}) or (\ref{12}). In the weak-field approximation the spherically symmetric in the Jordan frame is as
\begin{equation}\label{18}
ds^2=[1+2a(r)]dt^2-[1+2b(r)]^{-1}dr^2-r^2d\Omega.
\end{equation}
Where $a(r)$ and $b(r)$ are the functions of $r$. It was shown in (\cite{19}) that under the chameleon mechanism the post-Newton parameter, $\gamma_{PPN}=\frac{ b(r)}{a(r)}$, is approximately given by
\begin{equation}\label{19}
\gamma_{PPN}=\frac{3-\frac{\Delta r_c}{r_c}}{3+\frac{\Delta r_c}{r_c}}\simeq 1-\frac{2}{3}\frac{\Delta r_c}{r_c},
\end{equation}
presuming that the condition $m_{out}r\ll 1$ holds on the solar-system.
At the moment, if we apply  (\ref{17}) on the Earth and obtain the condition the thin-shell on the it, then  to presumption  that the Earth is a solid sphere with radius
 $R_e=6.37\times 10^3km$ and a mean density $\rho_e=5.52g/cm^3$  also is surrounded by an atmosphere with homogenous density $\rho_{atm}\simeq10^{-3}g/cm^3$ and $r_{atm}\simeq r_e$ as well as with the aid of the present tightest constraint on $\gamma_{PPN}$,
$\vert \gamma_{PPN}-1\vert<2.3\times 10^{-5}$ (\cite{20}), we obtain
 \begin{equation}\label{20}
\frac{\Delta r_c}{r_c}<3.45\times 10^{-5}.
\end{equation}
By substituting (\ref{17}) into (\ref{20}) and replace the dimensionless Earth potential $\Phi_c=\Phi_e=\frac{M_eG}{R_ec^2}=6.95\times 10^{-10}$ (\cite{34}),
\begin{equation}\label{21}
\epsilon<4.79\times 10^{-14}\frac{R_c}{\rho_{in}}
\end{equation}
By set $\rho_{in}=\rho_e=\frac{6\Phi_e}{r^2_e}=1.02\times 10^{-26}{cm}^{-2}$,  the equation (\ref{21}) gives $\epsilon<4.79\times10^{12}R_c$, so taking $Rc\sim \rho_{in}$, we have $\epsilon<4.79\times 10^{-14}$.
 It is observed that the deviation from the general relativity is very small.
It is notable that this model of $f(R)$ gravity has been studied in (\cite{35}), for a weak field limit.
In this regard, the line element  was  obtained as for $f(R)=R(\frac{R}{R_c})^{\mp \epsilon}$  model from $f(R)$ gravity  (\cite{1}), given as
\begin{equation}\label{211}
ds^2= (1-\frac{2M}{r}\mp2\epsilon \ln(r))dt^2-\frac{dr^2}{(1-\frac{2M}{r}\mp2\epsilon)}-r^2d\Omega
\end{equation}
This shows that
\begin{eqnarray}
a&=&-\frac{M}{r}\mp \epsilon \ln (r),\label{212}\\
b&=&-\frac{M}{r}\mp \epsilon. \label{213}
\end{eqnarray}
One can obtain  $\gamma_{PPN},$ up to second order of $\epsilon$ as
\begin{equation}\label{124}
\gamma_{PPN}=\frac{b}{a}\simeq1 \pm\frac{\epsilon r}{M}[1-\ln(r)]+{\cal O}({\epsilon}^2).
\end{equation}
Whereas, the most distance in the  solar system is around $5.5\times 10^{12}m,$  so that $\frac{M}{r}=\frac{GM}{rc^2}\simeq 2.5\times 10^{-8}$. Therefore one can obtain $\gamma_{PPN}\simeq1\mp 1.13\times 10^{-5}.$
This measure of $\gamma_{PPN}$, was obtained in the Jordan frame of $f(R)$ gravity is agreement  with
the  experimental consequence of $\gamma_{PPN}.$
\section{Equivalence Principle}
Now, we will consider experimental bounds  from a probable violation of the equivalence principle(EP).
For this doing, we suppose the Earth and its atmosphere to be an isolated body  far away from the effect of the other objets such as Moon, the Sun and the next planets. The same as the previous section that specified, the Earth has a mean density $\rho_e\simeq5.5 g/{cm}^3$ with radius $r_e=6.37\times 10^3km$ and the atmosphere density $\rho_{atm}\simeq 10^{-3}g/{cm}^3$. The region outside the atmosphere($r>r_{atm}$) has a homogeneous density $\rho_G\simeq10^{-24}g/{cm}^3$ (\cite{10}). Clearly, we know the gravitational potentials inside a spherically symmetric body, such as the Earth and the atmosphere as $\Phi_e\propto \rho_e r^2_e$, $\Phi_{atm}\propto \rho_{atm}r^2_{atm}$ respectively. As a result, we have  $\Phi_e \simeq 5.5\times 10^3\Phi_{atm}$  with assume $r_{atm}\simeq r_e $. From Eq. (\ref{15}), for the field values $\Phi_{atm}$ and $\Phi_G$ correspond with the regions $r_e<r<r_{atm}$ and $r>r_{atm} $ at their local minima of the effective potential respectively we have $\frac{\Delta r_{atm}}{r_{atm}}=\frac{(\Phi_G-\Phi_{atm})}{6\beta\Phi_{atm}}<1 .5\times 10^{-2 }$. Where $\Delta r_{atm}=10^2km$,  i.e.  (when the atmosphere has a thin-shell then the thickness of the shell ($\Delta r_{atm}$ ) is smaller than  the atmosphere $r_s=10-10^2km$) and $r_{atm}=6.5\times 10^3km$. Therefore, $\frac{\Delta r_e}{r_e}=\frac{(\Phi_{atm}-\Phi_{e})}{6\beta\phi_e}\simeq 2\times 10^{-4}\frac{\Delta r_{atm}}{r_{atm}}$. Where, we have  used from $\Phi_e=5.5\times10^{3}\Phi_{atm}$. As a result, the condition have a thin-shell for the atmosphere as
\begin{equation}\label{22}
\frac{\Delta r_e}{r_e}<3\times 10^{-6}.
\end{equation}
 By  making  use of the EP test, we want to measure the difference of  the free-fall acceleration of the Moon and the Earth toward the Sun. The constraint on the difference of two accelerations is given by (\cite{10})
 \begin{equation}\label{23}
\eta\equiv2\frac{\mid a_{Moon}-a_e\mid}{a_{Moon}+a_e}<10^{-13}.
\end{equation}
The acceleration induced by fifth force with the field profile $\phi(r)$ and the effective coupling $\beta_{eff}$
is  $ a^{fifth}=\mid\beta_{eff}\phi(r)\mid$. Then the acceleration $a_e$ and $a_{Moon}$ are (\cite{Khoury})
\begin{eqnarray}
a_e&\simeq&\frac{GM_\odot}{r^2}\left[ 1+3{(\frac{\Delta r_e}{r_e})}^2\frac{\Phi_e}{\Phi_\odot}\right] \label{24}\\
a_{Moon}&\simeq&\frac{GM_\odot}{r^2}\left[ 1+3{(\frac{\Delta r_e}{r_e})}^2\frac{{\Phi_e}^2}{\Phi_\odot\Phi_{Moon}}\right] \label{25}
\end{eqnarray}
Substituting the dimensionless potentials, $\Phi_\odot\simeq 2.1\times 10^{-6},\Phi_e\simeq6.95\times 10^{-10}$
 and $\Phi_{Moon}\simeq 3.1\times 10^{-11}$ into the Eq (\ref{24}, \ref{25}) and combine those with Eq (\ref{23}) gives
\begin{eqnarray}
\frac{\Delta r_e}{r_e}<2.1\times10^{-6}.\label{26}
\end{eqnarray}
Which is the same order of the condition (\ref{22}) as in the thin-shell condition for the atmosphere. Taking Eq. (\ref{26}) as the constraint of violation of EP and combining with (\ref{17}) one  can obtain
\begin{eqnarray}
\epsilon<\frac{0.3R_c}{\rho_{in}}\times10^{-14},\label{27}
\end{eqnarray}
which is not different from (\ref{21})
\section{Fifth Force}
We want to  consider the fifth force mediated by $\phi$. The dynamics of $\phi$  is still governed by the effective potential which is important when there is a component of matter whose energy-momentum tensor has nonzero trace. The solar system constrains get simply by making  the large field mass associated with the field $\phi$, is given by  $m^2_{min}=V_{eff,\phi\phi}(\phi_{min})$. The profile for a  potential connected with a fifth force interaction is given by a Yukawa potential between two tests masses $m_1$ and $m_2$ by distance $r$
as
\begin{equation}\label{28}
V(r)=-\alpha\frac{m_1m_2}{8\pi}\frac
{e^{-m_{\phi}r}}{r}
\end{equation}
Where $\alpha$ is strength of the interaction and $m_{\phi}^{-1}$ is the range. Therefore the fifth force experiment constrains regions of ( $\epsilon,m_{\phi}^{-1}$) parameter space. The experiments are
generally fulfilled in a vacuum chamber that the range of the interaction is  the same order  of  the spatial dimension of the chamber, namely  ${m_{\phi}}^{-1}\sim R_{vac}$. The tightest bound of the strength of the interaction is $\alpha<10^{-3} ($\cite{35}). We consider two identical bodies  with radius,  mass, uniform density $r_c,m_c,\rho_c$, respectively, in to the chamber. Assuming the thin-shell condition is satisfied by the two bodies, their field profile out side the bodies are given by (\cite{26})
\begin{equation}\label{29}
\phi(r)=-\frac{\beta}{4\pi}\frac{3\Delta r_c}{r_c}\frac{m_ce^{-\frac
{r}{R_{vac}}}}{r}+\phi_{vac}.
\end{equation}
Hence the corresponding potential energy of the interaction is
\begin{equation}\label{30}
V(r)=-2\beta ^2{(\frac{3\Delta r_c}{r_c})}^2\frac{{m_c}^2}{8\pi}\frac{e^{-r/R_{vac}}}{r}.
\end{equation}
The bound on the strength of the interaction to be given as follows
\begin{equation}\label{31}
2{\beta}^2{(\frac{3\Delta r_c}{r_c})}^2<10^{-3}.
\end{equation}
Writing Eq. (\ref{17}) for every one of the test bodies, we get
\begin{equation}\label{32}
\frac{\Delta r_c}{r_c}\approx[ \frac{\rho_{c}-\rho_{vac}}{12{\beta}^2\phi_c R_c}]\epsilon
\end{equation}
Here $\rho_{vac}$ is energy density of the vacuum inside the chamber. In the experiment performed in (\cite{35}), for a typical test body with mass $m_c\approx 40gr$ and radius $r_c\approx 1cm$ the density $\rho_c$ and the potential $\Phi_c$ have been obtained $9.5gr/{cm}^3,\, 10^{-27}$ respectively. Furthermore, the pressure in the vacuum chamber have been reported $3\times 10^{-8}mmHg$ which is equivalent to $\rho_{vac}\approx4.8\times 10^{-14}gr/{cm}^3$. Substituting the upon values in to (\ref{32}) and combining with (\ref{31}) result in  the bound
\begin{equation}\label{33}
\epsilon <3.6\times 10^{-29}\frac{R_c}{\rho_c}.
\end{equation}
Which if we replace  $\rho_c=1.7\times 10^{-26}{cm}^{-2}$, with substitute in to Eq. (\ref{33}) we can obtain $\epsilon<2\times10^{-3}R_c $
\section{Conclusion}
In this paper we have constrained $f(R)$ model of gravities, using the renowned equivalence between these models and scaler tensor theories.   Mostly, in this representation from $f(R)$ theory there is a strong coupling of the scalar field with the matter sector. We have used from the chameleon mechanism to suppress this coupling. We have shown that in order to the model
(\ref{11}) be consistent with the present local gravity experiments, the parameters $\epsilon$ and $R_c$ should satisfy the following conditions.
\begin{itemize}
\item The thin-shell condition of chameleon mechanism for model (\ref{11}), is satisfied for $\epsilon\lesssim 4.79\times 10^{-14}\frac{R_c}{\rho_e}$
\item This model is consistent with EP experiment for $\epsilon\lesssim 0.3 \times 10^{-14}\frac{R_c}{\rho_e}$
\end{itemize}
Whereas, the structure of this paper is perturbation state of general relativity, hence $\epsilon$ should be small. Therefore, we suggest $R_c$
must be  the same order of $\rho_e = 10^{-26} cm^{-2}$, where with this value of $R_c$, we have found that the model (\ref{11}) is consistent with the thin-shell and EP conditions for  $\epsilon\lesssim 10^{-14}$, as well as it is consistent with fifth force.
 At the moment, as a result applying  the above values from $\epsilon$, we obtain  $\gamma_{PPN}\simeq1\mp1.13\times 10^{-5}$. That is remarkable that this value
 of $\gamma_{PPN}$ are nearly equal to what is required by observation.
  It should be implied that the above result is achieved under the assumption that
$\rho_{out}\ll\rho_{in}\ll \frac{\Phi_c R_c}{\epsilon}$.  Therefore, the studied model is  a viable $f(R)$ model which satisfied EP and Solar System bounds and it has very small  deviation from the general relativity.

\end{document}